\documentclass[a4paper,11pt,onecolumn,accepted=2026-05-26]{quantumarticle}
\pdfoutput=1

\usepackage[utf8]{inputenc}
\usepackage[english]{babel}
\usepackage[T1]{fontenc}
\usepackage{amsmath,amssymb,mathtools}
\usepackage{hyperref}
\usepackage{physics}
\usepackage{booktabs}

\usepackage[numbers]{natbib}

\renewcommand{\emph}{\textit}

\newcommand{\aliceobs}[1]{A_{#1}}
\newcommand{\aliceproj}[2]{\Lambda^{#1}_{#2}}

\newcommand{\alicepovm}[2]{N^{#1}_{#2}}
\newcommand{\bobobs}[1]{B_{#1}}
\newcommand{\bobproj}[2]{\Pi^{#1}_{#2}}
\newcommand{\bobpovm}[2]{F^{#1}_{#2}}
\newcommand{\bobseq}[2]{{\mathsf B}^{#1}_{#2}}

\newcommand{\bobkraus}[2]{K^{#1}_{#2}}
\newcommand{\uli}{\overset{uli}{=}}
\newcommand{\til}{~}

\begin{document}
\title{Device-independent secure correlations in sequential quantum scenarios}

\author{Matteo Padovan}
\email{matteo.padovan.2@unipd.it}
\affiliation{Dipartimento di Ingegneria dell'Informazione, Universit\`a degli Studi di Padova, via Gradenigo 6B, IT-35131 Padova, Italy}
\orcid{0000-0001-9565-9438}

\author{Alessandro Rezzi}
\affiliation{Dipartimento di Ingegneria dell'Informazione, Universit\`a degli Studi di Padova, via Gradenigo 6B, IT-35131 Padova, Italy}

\author{Lorenzo Coccia}
\affiliation{Dipartimento di Ingegneria dell'Informazione, Universit\`a degli Studi di Padova, via Gradenigo 6B, IT-35131 Padova, Italy}
\orcid{0000-0003-1444-2355}

\begin{abstract}
    Device-independent quantum information is attracting significant attention, particularly for its applications in information security. 
    This interest arises from the fact that the security of device-independent protocols does not depend on the internal workings of the devices, but rather on the observed outcomes of spatially separated measurements together with the validity of quantum theory.
    Sequential scenarios, i.e., where measurements occur in a precise temporal order, have been proved to enhance performance of device-independent protocols in some specific cases by enabling the reuse of the same quantum state. 
    In this work, we propose a systematic approach to designing sequential quantum protocols for device-independent security.
    Our method begins with a bipartite self-testing qubit protocol and transforms it into a sequential protocol by replacing one measurement with a non-projective Positive Operator Valued Measurement (POVM) and adding an additional user thereafter. 
    We analytically prove that, with this systematic construction, the resulting ideal correlations are secure in the sense that they cannot be reproduced as a statistical mixture of other correlations, thereby enabling, for example, the device-independent certification of all the randomness present in the observed correlations.
    The general recipe we provide can be exploited for further development of new device-independent quantum schemes for security.
\end{abstract}

\maketitle

\section{Introduction}
The development of device-independent quantum information protocols is particularly appealing for security applications. 
The core idea behind this approach is to perform quantum information tasks and ensure their secure implementation based on observed data. 
Rather than relying on assumptions about the inner workings of the devices, security in this paradigm stems from the assumption that quantum mechanics can describe the correlations observed between two spatially separated users \cite{scarani2012device,Supic2020, Zapatero2023,Primaatmaja2023securityofdevice}. 
By removing the need for full experimental device characterization, device-independent protocols not only strengthen security but can also simplify its certification in practical implementations.

Device-independent protocols rely on quantum nonlocality \cite{Brunner2014,
Christensen2015,Goh_2018, Acin2012, Miller_2017}, i.e., the fact that quantum correlations cannot be explained by any local hidden variables model. 
This is achieved by violating a Bell inequality, and since Bell inequalities are linear functions of only the observed correlations, this certification is possible in a device-independent setting. 
Crucially, such certification requires the shared quantum system to be in an entangled state, as entanglement is necessary for nonlocal correlations to arise \cite{Brunner2014}.
However, entanglement is a delicate resource \cite{HorodeckiEntanglementRev}: creating and maintaining high-quality entangled states is technically demanding. 
Additionally, entanglement is completely lost\footnote{In some common implementations, not only the entanglement but the physical state itself is lost during measurement, for example, in photonic realizations where photons are absorbed to measure their properties. We discuss this aspect further in the Conclusions.} when observers use rank-one projective measurements, which is the default choice in most current device-independent protocols.

Non-projective measurements offer a way to circumvent this last fundamental limitation. 
These measurements allow the entanglement to be preserved in the post-measurement state \cite{Silva2015,Foletto2020}, and this feature can be leveraged for sequential measurement schemes with the aim of improving the performances of quantum protocols. 
For example, sequential measurements can be exploited in quantum networks for sharing nonlocality \cite{Zhang2023}, or utilized for randomness generation, enhancing the number of random bits produced in each round \cite{Curchod2017, Foletto2021, Padovan2024}. 
However, to date, the literature lacks simple sequential schemes proved to be robust against noise, or a systematic way of constructing sequential protocols.

To address this last point we take inspiration  from \cite{Padovan2024}, where the authors introduced a sequential quantum protocol that can be used for randomness generation.
They analytically computed the quantum min-entropy for a specific sequence of measurements under ideal conditions and provided numerical analysis of the performance in presence of depolarizing noise.
In this work, we generalize such scenario into a broader class of sequential quantum protocols suitable for device-independent randomness generation. 
We consider a general bipartite self-testing protocol with real observables and a maximally entangled qubit state.
We propose a systematic sequential extension of it by changing a measurement into a non-projective one and adding further sequential measurements after it.
The entire protocol involves three parties: two sequential parties, Bob$_1$ and Bob$_2$, and a spatially separated party, Alice, each of them with a binary choice of measurement with dichotomic outcomes.

We demonstrate, under device-independent assumptions, that the ideal quantum correlations generated by our sequential extension cannot be replicated by statistically selecting alternative strategies which could benefit a potential eavesdropper. 
In technical terms, this means that the correlations are extremal, ensuring that no eavesdropper can gain any advantage beyond simply betting on the most probable outcome.
In fact, the extremality condition also implies that the quantum min-entropy and the worst-case conditional von Neumann entropy of the measurement outcomes reduce, respectively, to the classical (non-conditional) min-entropy and Shannon entropy of the outcomes probability distribution. 
Therefore we can certify all the randomness obtainable from the observed outcomes.

\section{Methods}
\subsection{Sequential quantum correlations}
\label{sec:seq_quant_corrs}
The scenario we consider involves three users: Alice, Bob$_1$, and Bob$_2$, each with a device that takes an input and produces an output. 
Alice’s inputs are labeled $x \in \qty{0,1}$, with outcomes $a \in \qty{\pm 1}$, while Bob$_n$’s inputs are denoted by $y_n \in \qty{0,1}$, with corresponding outcomes $b_n \in \qty{\pm 1}$.

\begin{figure}
    \centering
    \includegraphics[width=.6\linewidth]{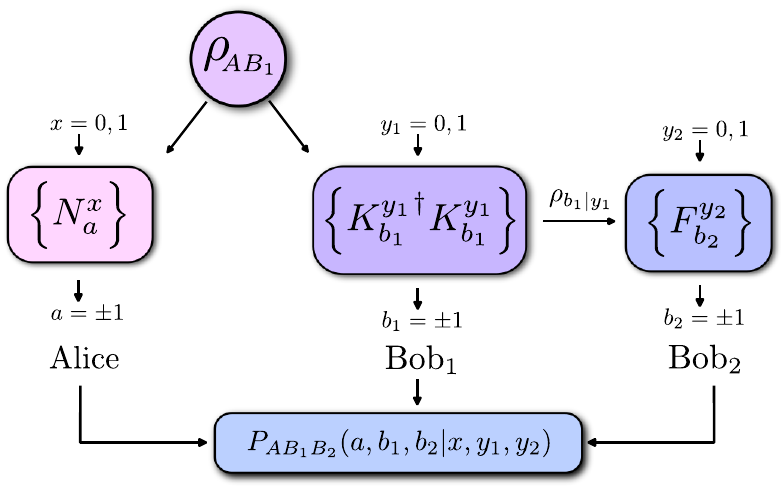}
    \caption{The sequential quantum scenario considered in this work.}
    \label{fig:boxes}
\end{figure}

The scheme proceeds as follows: a source prepares an unknown physical state $\rho_{AB_1}$, which is sent to both Alice and Bob$_1$.
They locally randomly choose an input for their own device, which returns the outcome according to the result of a Positive Operator-Valued Measurement (POVM).
After his operations, Bob$_1$ sends its post-measurement state to a second Bob, Bob$_2$, who also measures the state.
We label Alice's POVMs as $\{\alicepovm{x}{a}\}$, Bob$_1$'s POVMs as $\{{\bobkraus{y_1}{b_1}}^\dagger \bobkraus{y_1}{b_1}\}$, for some appropriate set of Kraus operators, and Bob$_2$'s POVMs with $\{\bobpovm{y_2}{b_2}\}$.
We model the POVMs of Bob$_1$ with Kraus operators because we are interested in the post-measurement state left to Bob$_2$ which, given the input $y_1$, is
\begin{equation}
\label{eq:cptp_map}
    \rho_{AB_2}^{y_1}\equiv\sum_{b_1} \bobkraus{y_1}{b_1}\rho_{AB_1}{\bobkraus{y_1}{b_1}}^{\dagger} \,.
\end{equation}
Throughout this process, all inputs are assumed to be independent, and there is no communication between Alice and the Bobs. 
However, Bob$_1$ is allowed to transmit information to Bob$_2$ after his measurement and, without loss of generality, this information is assumed to be passed through the action of the Kraus operators \cite{Bowles2020}.

The conditional sequential correlations between the three users, $P_{AB_1B_2}(a,\mathbf{b}|x,\mathbf{y})$ with $\mathbf{b}=(b_1,b_2)$ and $\mathbf{y}=(y_1,y_2)$, are described through the Born rule:
\begin{equation}
\label{eq:povm_correlations}
    P_{AB_1B_2}(a,\mathbf{b}|x,\mathbf{y}) = \Tr\left[ \rho_{AB_1} \til\alicepovm{x}{a} \otimes {\bobkraus{y_1}{b_1}}^\dagger \bobpovm{y_2}{b_2} \bobkraus{y_1}{b_1} \right]\,.
\end{equation}
From the above joint correlations, the marginals can be calculated by summing over the other parties' outcomes.
In particular, we can calculate the marginal correlations between Alice-Bob$_1$ and Alice-Bob$_2$ as
\begin{align}
    P_{AB_1}(a,b_1|x,y_1) &= \Tr\left[ \rho_{AB_1} \til \alicepovm{x}{a} \otimes {\bobkraus{y_1}{b_1}}^\dagger \bobkraus{y_1}{b_1}\right]\,,\\
    P_{AB_2}(a,b_2|x,y_1,y_2) &= \Tr\left[ \rho_{AB_1}\til \alicepovm{x}{a} \otimes \sum_{b_1} {\bobkraus{y_1}{b_1}}^\dagger \bobpovm{y_2}{b_2} \bobkraus{y_1}{b_1}\right] \, \label{eq:AB2_marg_POVM}.
\end{align}
From now on, we assume no finite statistics effect and that the correlations arise from independent and identically distributed (i.i.d.) runs of the protocol.

\subsection{A class of sequential quantum protocols}
\label{sec:seq_prot}
\begin{figure}
    \centering
    \includegraphics[width=.9\linewidth]{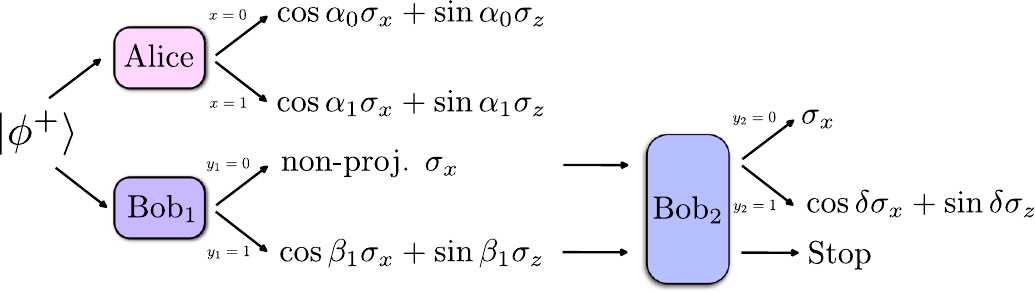}
    \caption{Scheme of measurement of the ideal sequential protocols. Alice chooses between two dichotomic measurements. Bob$_1$ chooses between a non-projective measurement ($y_1 = 0$) or a projective one ($y_1=1$).
    The first one is defined by Kraus operators \eqref{eq:Kpiu} and \eqref{eq:Kmeno} parametrized by a strength parameter $\theta$: when $\theta=0$ or $\theta=\pi/4$, the Kraus operators reduces to the eigen-projectors of $\sigma_x$ or to the rescaled identity. Bob$_2$ receives the post-measurement state and, depending on the input of Bob$_1$ measures one of two possible projective measurements (if $y_0 = 0$) or stops the run of the protocol (if $y_1=1$).  }
    \label{fig:meas}
\end{figure}

Having outlined the sequential quantum scenario, we now present a class of sequential protocols for device-independent randomness generation. 
These protocols can be understood as sequential extensions of standard bipartite protocols that admit a self-testing certification from the correlations they produce.
Therefore, we begin by considering a general self-testing qubit protocol with maximally entangled states, and then we present our systematic sequential construction.

When Alice and Bob share the two-qubit maximally entangled state 
\begin{equation}
\label{eq:state}
    \ket*{\phi^+} = \frac{\ket{00}+\ket{11}}{\sqrt{2}}\,,
\end{equation}
the most general measurements that can be chosen to realize self-testing (in the two-inputs and two-outputs scenario) are described by the observables
\begin{align}
    \mathcal{A}_0 &= \cos \alpha_0 \sigma_x + \sin \alpha_0 \sigma_z \label{eq:A0} \ , \\ 
    \mathcal{A}_1  &= \cos \alpha_1 \sigma_x + \sin \alpha_1 \sigma_z \label{eq:A1} \ , \\
    \mathcal{B}_0 ' &= \sigma_x \label{eq:B0} \ , \\
    \mathcal{B}_1 &= \cos \beta_1 \sigma_x + \sin \beta_1 \sigma_z \label{eq:B1}\,,
\end{align}
with the parameters $\alpha_0$, $\alpha_1$, and $\beta_1$ chosen to satisfy the necessary and sufficient conditions discussed in \cite{Wang_2016}.
We only consider real measurements: this stems from the fact that, once the necessary and sufficient conditions for the self-testing are met, we can always find local isometries mapping the measurements of Alice and Bob to combinations of $\sigma_x$ and $\sigma_z$ \cite{Wang_2016,McKague2012}.
In principle, we could have also defined $\mathcal{B}_0' = \cos \beta_0 \sigma_x + \sin \beta_0 \sigma_z$. 
However, since the correlations $\expval{\mathcal{A}_x\mathcal{B}_y}=\cos(\alpha_x-\beta_y)$, when evaluated on the state $\ket{\phi^+}$, depend only on the difference $\alpha_x-\beta_y$, we do not lose generality by assuming the angle $\beta_0=0$.

When extending the protocol in \eqref{eq:state} and \eqref{eq:A0}-\eqref{eq:B1} sequentially, we aim to use the same state and modify one measurement into a non-projective one to prevent the loss of entanglement after it is performed.
At the same time we want to preserve the self-testing properties to facilitate security proofs. 
To achieve this, we replace $\mathcal{B}_0'$ by a POVM $\{{\bobkraus{0}{b_1}}^\dagger\bobkraus{0}{b_1}\}$ described by the Kraus operators
\begin{align}
    \bobkraus{0}{+} &\equiv \cos\theta \frac{\openone + \mathcal{B}_0'}{2} + \sin\theta \frac{\openone - \mathcal{B}_0'}{2}\,, \label{eq:Kpiu}\\
    \bobkraus{0}{-} &\equiv \sin\theta \frac{\openone + \mathcal{B}_0'}{2} + \cos\theta \frac{\openone - \mathcal{B}_0'}{2}\,, \label{eq:Kmeno}
\end{align}
where we call $\theta\in\qty[0,\frac{\pi}{4}]$ the strength parameter and $(\openone \pm \mathcal{B}_0')/2$ define the two eigenspace projectors of $\mathcal{B}_0'$.
Note that for $\theta = 0$ the two Kraus operators reduce to the two projectors of $\mathcal{B}_0'$, while for $\theta = \frac{\pi}{4}$ they become proportional to the identity operator. 
After this measurement, the post-measurement state is sent to a second Bob, who performs a projective measurement equal to $\mathcal{B}_0'$ for $y_2=0$ and a projective measurement in the form $\cos \delta \sigma_x + \sin \delta \sigma_z$, with a free parameter $\delta \neq 0$,  for $y_2=1$. 

The key point is that, for the choice $y_1=0$ and $y_2=0$, Bob$_2$ measures $\mathcal{B}_0'$ after Bob$_1$ has performed a non-projective version of $\mathcal{B}_0'$ itself.
This implies that the correlations $ P_{AB_2}(a,b_2|x,0,0) $ between Alice and Bob$_2$ are equivalent to those that would be obtained with a projective measurement of $\mathcal{B}_0'$, i.e., those which allow the self-testing.
To check this property, we evaluate \eqref{eq:AB2_marg_POVM} for $y_1=0$ and $y_2=0$:
\begin{equation}
\begin{split}
    P_{AB_2}(a,b_2|x,0,0) &= \sum_{b_1} P_{AB_1B_2}(a,b_1,b_2|x,0,0) \\
    &= \sum_{b_1} \Tr\qty[ \rho_{AB_1} \til\alicepovm{x}{a} \otimes {\bobkraus{0}{b_1}}^\dagger \frac{\openone+b_2 \mathcal{B}_0'}{2}\bobkraus{0}{b_1}] \\
    &= \Tr\left[ \rho_{AB_1} \til\alicepovm{x}{a} \otimes \frac{\openone+b_2 \mathcal{B}_0'}{2}\right]\,.
\end{split}
\end{equation}
Last equality holds because the projectors of $\mathcal{B}_0'$ are invariant under the action of the Kraus operators $\bobkraus{0}{b_1}$:
\begin{equation}
\label{eq:krausB0nomod}
    \sum_{b_1} \bobkraus{0}{b_1} \frac{\openone + b_2 \mathcal{B}_0'}{2} \bobkraus{0}{b_1} = \frac{\openone + b_2 \mathcal{B}_0'}{2}\,.
\end{equation}
Thanks to this property, after the sequential extension, we can still rely on the self-testing properties of the correlations, which we will leverage to prove randomness in a device-independent scenario.

The ideal sequential protocol, depicted in Fig.\ \ref{fig:meas}, can be summarized as follows.
First, Alice and Bob$_1$ share a maximally entangled state, $\ket{\phi^+}_{AB_1}$. Alice randomly selects one of two inputs, $x \in \{0,1\}$, corresponding to the two observables $\mathcal{A}_x$ defined in \eqref{eq:A0} and \eqref{eq:A1}.
Bob$_1$ also randomly chooses between two inputs, $y_1 \in \{0,1\}$. Input $y_1 = 1$ corresponds to a projective measurement of $\mathcal{B}_1$, while $y_1 = 0$ corresponds to a non-projective measurement of $\mathcal{B}_0'$, implemented through the two Kraus operators \eqref{eq:Kpiu} and \eqref{eq:Kmeno}.
If $y_1 = 0$, Bob$_1$ sends its post-measurement state to Bob$_2$. 
Bob$_2$ then randomly selects one of two inputs, $y_2 \in \{0,1\}$, corresponding to either a projective measurement of $\mathcal{B}_0'$ or a measurement of the observable $\cos \delta \, \sigma_x + \sin \delta \, \sigma_z$.
Otherwise, if Bob$_1$ chooses $y_1 = 1$, then Bob$_2$ does nothing, ending the protocol. 

\subsection{Device-independent framework for sequential quantum correlations}
\label{sec:dilation}
We now recall the projective framework introduced in \cite{Padovan2024} for describing sequential quantum correlations in a device-independent scenario.
We adopt this framework for our analytical proofs.

In a device-independent sequential scenario, the experimental correlations \eqref{eq:povm_correlations} are assumed to derive from an unknown tripartite state $\ket{\psi} \in \mathcal{H}_A\otimes\mathcal{H}_B\otimes\mathcal{H}_E$ measured by  Alice's and Bob's unknown projectors acting on the Hilbert spaces $\mathcal{H}_A$ and $\mathcal{H}_B$ respectively. $\mathcal{H}_E$ is an arbitrary Hilbert space on which a potential eavesdropper Eve acts.
Therefore, the sequential correlations can be calculated as 
\begin{equation}
\label{eq:proj_correlations}
    P_{AB_1B_2}(a,\mathbf{b}|x,\mathbf{y}) = \expval{\aliceproj{x}{a}\otimes \bobproj{y_1}{b_1}\bobproj{y_1y_2}{b_2}\otimes \openone_E}{\psi}\,.
\end{equation}
In this framework, Alice's POVMs are replaced by the projective measurements $\{\aliceproj{x}{a}\}_x$, and the pair of Bobs, represented by the four-outcomes POVMs $\{{\bobkraus{y_1}{b_1}}^\dagger \bobpovm{y_2}{b_2} \bobkraus{y_1}{b_1}\}_{b_1,b_2}$ in \eqref{eq:povm_correlations}, is replaced by just one Bob with the product of the projectors $\bobproj{y_1}{b_1}$ and $\bobproj{y_1y_2}{b_2}$.
This product structure is derived from the Stinespring dilation of the four-outcome POVMs.
In particular, by defining the four orthogonal projectors $\{\bobseq{y_1y_2}{b_1b_2}\}_{b_1,b_2}$ as the Stinespring dilation of $\{{\bobkraus{y_1}{b_1}}^\dagger \bobpovm{y_2}{b_2} \bobkraus{y_1}{b_1}\}_{b_1,b_2}$ such that
\begin{equation}
    P_{AB_1B_2}(a,\mathbf{b}|x,\mathbf{y}) = \expval{\aliceproj{x}{a}\otimes \bobseq{y_1y_2}{b_1b_2}\otimes \openone_E}{\psi}\,,
\end{equation}
then the projectors of the two Bobs, $\bobproj{y_1}{b_1}$ and $\bobproj{y_1y_2}{b_2}$, are defined as
\begin{align}
    \bobproj{y_1}{b_1} &\equiv \sum_{b_2} \bobseq{y_1y_2}{b_1b_2}\,, \label{eq:bob1marginalproj}\\
    \bobproj{y_1y_2}{b_2} &\equiv \sum_{b_1} \bobseq{y_1y_2}{b_1b_2} \,.
\end{align}
As discussed in \cite{Padovan2024}, sequentiality implies that the left-hand side of \eqref{eq:bob1marginalproj} is independent of $y_2$.
From these two equations we can deduce the commutation relation $\left[\bobproj{y_1}{b_1}, \bobproj{y_1y_2}{b_2}\right] = 0$ for any input pair $y_1, y_2$, 
and that the marginal distributions of Bob$_1$ and Bob$_2$ can written as:
\begin{align}
    P_{AB_1}(a,b_1|x,y_1) &= \expval{\aliceproj{x}{a}\otimes \bobproj{y_1}{b_1}\otimes \openone_E}{\psi} \,, \\
    P_{AB_2}(a,b_2|x,y_1,y_2) &= \expval{\aliceproj{x}{a}\otimes \bobproj{y_1y_2}{b_2}\otimes \openone_E}{\psi} \,.
\end{align}
From the orthogonality constraint $\bobseq{y_1y_2}{b_1b_2}\bobseq{y_1y_2}{b_1'b_2'} = \delta_{b_1,b_1'}\delta_{b_2,b_2'} \bobseq{y_1y_2}{b_1b_2}$, we can see that $\bobseq{y_1y_2}{b_1b_2} = \bobproj{y_1}{b_1}\bobproj{y_1y_2}{b_2}$, therefore such product is a well-defined measurable operator defining the joint sequential correlations of the two Bobs.
In addition, since every measurement of the protocol has two possible outcomes, we can construct Hermitian and unitary operators from the projectors in \eqref{eq:proj_correlations} as:
\begin{align}
    \aliceobs{x} &\equiv \aliceproj{x}{+} - \aliceproj{x}{-} \label{eq:summary1}\, ,  \\
    \bobobs{y_1} &\equiv \bobproj{y_1}{+} - \bobproj{y_1}{-} \, ,  \\
    \bobobs{y_1y_2} &\equiv \bobproj{y_1y_2}{+} - \bobproj{y_1y_2}{-} \,.
\end{align}
Using these operators we can provide an alternative equivalent parametrization for the sequential correlations \eqref{eq:proj_correlations}.
In particular, the observables $B_1$ and $B_{00}$ generate the same correlations as $\mathcal{B}_1$ and $\mathcal{B}_{0}'$, and hence they can be self-tested.

\section{Results}
\subsection{Security and extremality of correlations}
In this section, we explain why the correlations produced by the sequential construction of Sec.\ \ref{sec:seq_prot} can be used for device-independent security.
By security, we mean that for any possible quantum realization of the sequential correlations, the following holds:
\begin{equation}
\label{eq:security_def}
    \expval{\aliceproj{x}{a}\otimes \bobproj{y_1}{b_1}\bobproj{y_1y_2}{b_2}\otimes E}{\psi} = P_{AB_1B_2}(a,\mathbf{b}|x,\mathbf{y})\expval{E}\,,
\end{equation}
where $E$ is an arbitrary Hermitian quantum operator measured by Eve. 
As shown in \cite{Franz2011}, this condition is equivalent to requiring that the correlations are extremal, meaning they cannot be expressed as a combination of other quantum correlations. 
Intuitively, this implies that an eavesdropper cannot decompose them to gain an advantage. See also Appendix \ref{app:extremality} for more details.

A necessary condition for the correlations to be extremal (and hence secure) is that they lie on the boundary of the set of all realizable quantum correlations \cite{Goh_2018}. 
In our scenario, we denote this set as $\mathcal{Q}_{SEQ}$, which includes all quantum correlations between one Alice and two Bobs that are compatible with the sequentiality constraints \cite{Bowles2020}.
The fact that some correlations lie on the boundary of $\mathcal{Q}_{SEQ}$ can be certified by demonstrating that they saturate a Tsirelson bound \cite{Cirelson1980quantum,Brunner2014, Bowles2020}. 
Indeed, relying on the results of \cite{coccia2025}, we can explicitly construct a Tsirelson bound that is saturated by the correlations produced by the sequential protocol of Sec.\ \ref{sec:seq_prot}, as we now show.

First, recall that we constructed our class of sequential protocols starting from a non-sequential class, \eqref{eq:state} and \eqref{eq:A0}--\eqref{eq:B1}.
The correlations produced by this class of protocols self-test state and measurements \cite{Wang_2016} and our construction of sequential protocols maintains this property, as discussed in Section \ref{sec:seq_prot}. 
More in detail, thanks to the invariance \eqref{eq:krausB0nomod}, the correlations Alice-Bob$_1$, $P_{AB_1}(a,b_1|x,1)$, and the correlations Alice-Bob$_2$, $P_{AB_2}(a,b_2|x,0,0)$ certify, up to local isometries, that the measurements $\aliceobs{0}$, $\aliceobs{1}$, $\bobobs{00}$ and $\bobobs{1}$ are the ones in \eqref{eq:A0}--\eqref{eq:B1}, and the part of the state on which they act is $\ket{\phi^+}$.

Regarding the correlations of the remaining operators, $\bobobs{0}$ and $\bobobs{01}$, we observe that they are realized by a rank-one four-outcomes POVM with elements $\{ {\bobkraus{0}{b_1}}^\dagger \bobpovm{1}{b_2} \bobkraus{0}{b_1}\}$. 
Then, using the results of \cite{coccia2025}, the Tsirelson bound we consider is written as
\begin{equation}
\label{eq:bound}
    \expval{S_{\theta,\delta} } = 0\,,
\end{equation}
where the Bell operator $S_{\theta,\delta}$ has the form
\begin{equation}
    S_{\theta,\delta} \equiv \frac{1}{2}\Bigl[\openone - X_A \Bigl( c_1\bobobs{01}+c_2\bobobs{0} \Bigr) - Z_A\Bigl( c_3\bobobs{01}+c_4\bobobs{0} \Bigr)\Bigr]\,.
    \label{eq:Std}
\end{equation}
The operators $X_A$ and $Z_A$ are defined from $\aliceobs{0}$ and $\aliceobs{1}$ such that in the ideal protocol they coincide with $\sigma_x$ and $\sigma_z$ respectively:
\begin{equation}
\begin{split}
    X_A &\equiv \frac{\sin\alpha_1 A_0+\sin\alpha_0 A_1}{\sin(\alpha_1-\alpha_0)} \ ,\\
    Z_A &\equiv \frac{\cos\alpha_1 A_0+\cos\alpha_0 A_1}{\sin(\alpha_0-\alpha_1)}\,,
\end{split}
\end{equation}
while the coefficients $c_i$ depend on the strength parameter $\theta$ and on the angle $\delta$ (see Eq.\ \eqref{eq:C_coefficients} in Appendix \ref{app:proofs}). 
Since it holds that $\expval{S_{\theta,\delta} } = \expval*{S_{\theta,\delta}^\dagger S_{\theta,\delta} }\geq 0$, then Eq.\ \eqref{eq:bound} implies 
\begin{equation}
    S_{\theta,\delta} \ket{\psi} = 0\,,
\end{equation}
from which the security condition can be verified.
All the calculations are provided in Appendix \ref{app:securitycalc}.

The proof holds for any value of the strength parameter $\theta$, except for the extremal values $0$ and $\frac{\pi}{4}$.
This can be explained by considering that when $\theta = 0$, Bob$_1$ is performing a projective measurement and hence Bob$_2$ receives a state that is no longer entangled.
In contrast, when $\theta = \frac{\pi}{4}$, although Bob$_1$ shares a maximally entangled state, his outcomes are completely uncorrelated with both Alice and Bob$_2$, because his measurement operators are proportional to the identity.
However, as shown by the following examples, setting $\theta \in \qty{0,\frac{\pi}{4}}$ does not necessarily imply that the randomness drops to zero, but rather to a value that depends on the specific sequential protocol chosen.

The security condition \eqref{eq:security_def} guarantees that the quantum min-entropy reduces to the classical min-entropy, and the worst-case conditional von Neumann entropy reduces to the Shannon entropy, as we show in Appendix \ref{app:extremality}.
These quantities are defined in the following way \cite{Nieto-Silleras_2014, Bancal_2014, Brown_2020}.
Suppose that we aim to generate random numbers from the outcomes obtained by the inputs $x^\star$ and $\mathbf{y}^\star$.
By using the projective framework of \eqref{eq:proj_correlations}, we define the classical-quantum post-measurement state associated to the measurements $x^\star$ and $\mathbf{y}^\star$ (not to be confused with the post-measurement state of Bob$_1$ introduced in Section \ref{sec:seq_quant_corrs}):
\begin{equation}
\label{eq:cq_pm_state}
    \rho_{ABE}^{post} \equiv \sum_{a,\mathbf{b}} P_{AB_1B_2}(a,\mathbf{b}|x^\star,\mathbf{y}^\star) \dyad{a,\mathbf{b}}\otimes \tau_E^{a,\mathbf{b}}\,.
\end{equation}
In the above definition, $\dyad{a,\mathbf{b}}\in\mathcal{H}_{A'}\otimes\mathcal{H}_{B'}$ is a classical state in which Alice and Bob register their measurement results, and $\tau_E^{a,\mathbf{b}}\in\mathcal{H}_E$ is the reduced state of Eve:
\begin{equation}
    \tau_E^{a,\mathbf{b}} \equiv \frac{\Tr_{AB}\qty[\aliceproj{x^\star}{a}\otimes\bobproj{y_1^\star}{b_1}\bobproj{y_1^\star,y_2^\star}{b_2}\otimes\openone_E\til\dyad{\psi}]}{P_{AB_1B_2}(a,\mathbf{b}|x^\star,\mathbf{y}^\star)}\,.
\end{equation}
The worst-case conditional von Neumann entropy is defined as
\begin{equation}
H(AB|E) \equiv \min_{\rho_{ABE}^{post}}\left[ H(AB)_{\rho_{ABE}^{post}} - H(E)_{\rho_{ABE}^{post}}\right] \ ,  
\end{equation}
where the minimization is over all the possible classical-quantum post-measurements states compatible with the observed probability distribution.
The min-entropy is defined as $H_{min} \equiv -\log_2 G(AB|E)$, where $G(AB|E)$ is the guessing probability 
\begin{equation}
    G(AB|E) \equiv \! \! \! \max_{E_{a\mathbf{b}},\tau_{E}^{a,\mathbf{b}}} \sum_{a,\mathbf{b}} \! P_{AB_1B_2}(a,\mathbf{b}|x^\star,\mathbf{y}^\star) \Tr_E\qty[E_{a\mathbf{b}} \tau_{E}^{a,\mathbf{b}}]
\end{equation}
where, again, the maximization must be compatible with the experimental data and $E_{a\mathbf{b}}$ is a generic projector of Eve.
It turns out that, if the security definition \eqref{eq:security_def} holds, then the post-measurement state \eqref{eq:cq_pm_state} has a product form $\rho_{ABE}^{post} = \rho_{AB}^{post}\otimes\rho_{E}^{post}$. 
Therefore, $H(AB|E)$ reduces to the Shannon entropy of $P_{AB_1B_2}(a,\mathbf{b}|x^\star,\mathbf{y}^\star)$, and the guessing probability to the best classical guess $\max_{a,\mathbf{b}} P_{AB_1B_2}(a,\mathbf{b}|x^\star,\mathbf{y}^\star)$.

In our protocol the input sequence $\mathbf{y} = (0,0)$ is used to generate the self-testing correlations, while the sequence $\mathbf{y}^\star = (0,1)$ is chosen for randomness generation. 
To maximize the entropy, $\delta$ must be chosen based on the values of $\alpha_0$ (if $x^\star = 0$) or $\alpha_1$ (if $x^\star = 1$). 
The complete parametric expressions for the min-entropy and von Neumann entropy are provided in Eq.\ \eqref{eq:entropies_protocols} in Appendix \ref{app:securitycalc}.

\subsection{Example 1: Sequential CHSH}
\begin{figure}
    \centering
    \includegraphics[width=.6\linewidth]{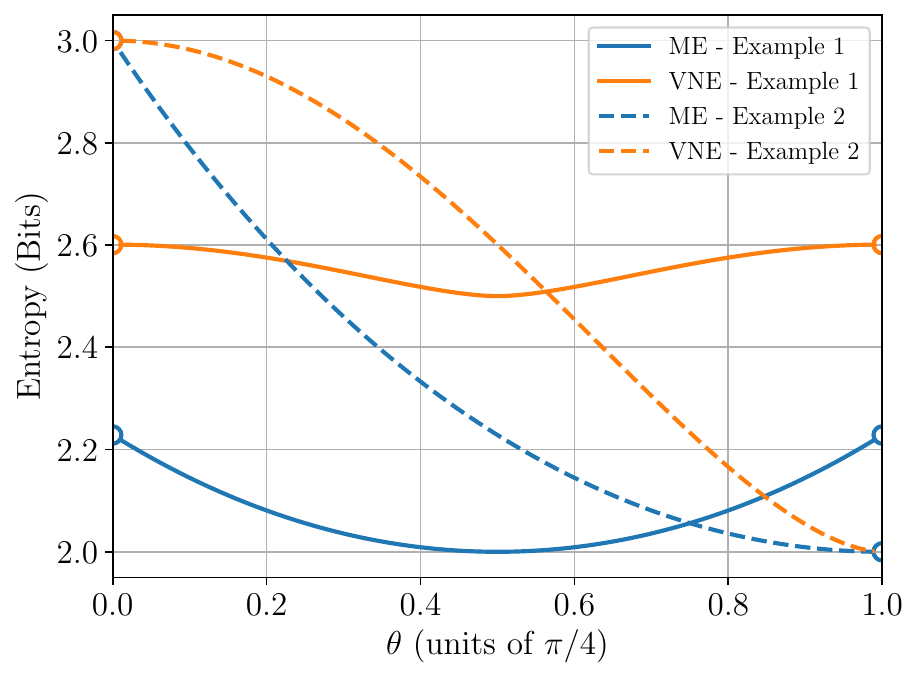}
    \caption{Min-entropy (ME) and von Neumann entropy (VNE) of the protocols described in Example 1 (continuous line) and Example 2 (dashed line). As discussed in the main text, when $\theta \in \{0,\pi/4\}$, the result \eqref{eq:security_def} does not hold anymore, shown by mean of empty half circles, and only one measurement of the Bobs' sequence can be certified.
At those points, the certified min (von Neumann) entropy is $\approx 1.2$ bits ($\approx 1.6$ bits) for Example 1. For Example 2, when $\theta=0$ both entropies reduce to 2 bits, while when $\theta=\pi/4$ they reduce to 1 bit.
The analytical expression of these entropies can be found in \eqref{eq:entropies_protocols}.}
    \label{fig:entropies}
\end{figure}

In \cite{Padovan2024}, the authors proposed a sequential extension of the CHSH protocol and analytically proved the local min-entropy of a specific Bobs' input sequence in case of ideal correlations.
With respect to the general class of protocols presented before, the sequential CHSH is defined by setting $\alpha_0 = \frac{\pi}{4}$, $\alpha_1 = \frac{3\pi}{4}$, $\beta_1 = \frac{\pi}{2}$ and $\delta = \frac{\pi}{2}$, and by performing an Hadamard transformation on Bobs' measurements.
We can conclude that whatever input (global or local) is chosen for randomness generation, under ideal conditions, the min-entropy coincides with the entropy of the most probable outcome, and the von Neumann entropy with the Shannon entropy of the outcomes probability distribution.
See Fig.\ \ref{fig:entropies} for a plots of the two entropies when the randomness is generated by the measurements identified by $x^\star=0$, $y_1^\star=0$ and $y_2^\star=1$.
For $\theta \in \qty{0,\frac{\pi}{4}}$ the min-entropy reduces to $\approx 1.2$ bits, while the von Neumann entropy reduces to $\approx 1.6$ bits.

\subsection{Example 2: Sequential extension of \cite{Wooltorton2022}}
The previous example is not optimal for global randomness generation because the ideal measurements of Alice and the Bobs are strongly correlated.
Therefore, we could look for a protocol in which Alice has at least one measurement that is loosely correlated with the Bobs' measurements.
One feasible candidate is the sequential extension of the protocol proposed in \cite{Wooltorton2022}, saturating the Bell expression
\begin{equation}
    \expval{I_{\omega}} = \expval{\aliceobs{0}\bobobs{0}} + \frac{\expval{\aliceobs{0}\bobobs{1}}+\expval{\aliceobs{1}\bobobs{0}}}{\sin\omega} - \frac{\expval{\aliceobs{1}\bobobs{1}}}{\cos2\omega}\,,
\end{equation}
for $\omega \in (0, \frac{\pi}{6}]$.
The Tsirelson bound is given by $2\cos^3\omega/(\cos2\omega\sin\omega)$, and the optimal non-sequential strategy is to set $\alpha_0 = \frac{\pi}{2}$, $\alpha_1 = -\omega$ and $\beta_1 = \omega + \frac{\pi}{2}$.
For the sequential extension, we set $\delta = \frac{\pi}{2}$, since, as can be verified, it maximizes the generation of randomness.

For a plot of the min- and von Neumann entropy generated by the measurements identified by $x^\star=0$, $y_1^\star=0$ and $y_2^\star=1$ with $\omega = \frac{\pi}{6}$, see Fig.\ \ref{fig:entropies}. 
It shows that both entropies are always greater than $2$ bits for $\theta \in (0,\frac{\pi}{4})$ and reduce to $2$ bits only for $\theta = 0$ and to 1 bit for $\theta=\pi/4$.

\begin{figure}
    \centering
    \includegraphics[width=.7\linewidth]{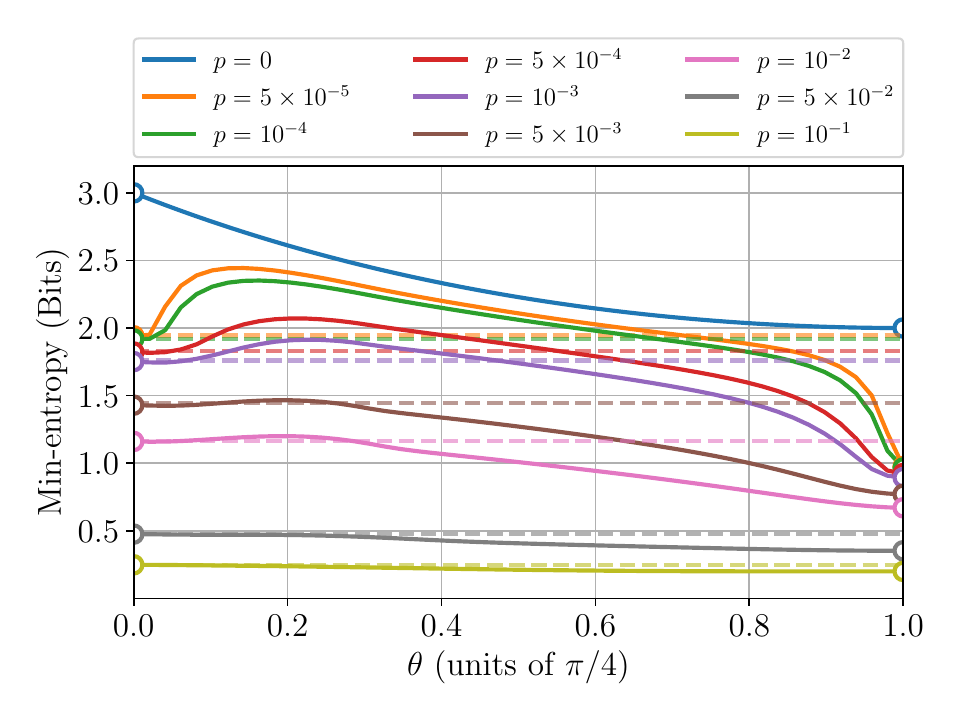}
    \caption{Lower bound on the min-entropy of the protocol described in Example 2 as a function of the strength parameter $\theta$ for several values of depolarizing noise $p$. These numerical simulations are performed by leveraging semidefinite programming (SDP) and the NPA hierarchy \cite{navascues_bounding_2007,navascues_convergent_2008} at order $2+AAB+ABB$ when the correlations $P^{exp}_{AB_1B_2} = (1-p) P^{ideal}_{AB_1B_2} + p P^{random}_{AB_1B_2}$ are observed. The term $P^{ideal}_{AB_1B_2}$ denotes the correlations obtained with ideal state and measurements and $P^{random}_{AB_1B_2} = 1/8$ are completely random correlations.
    The NPA order is chosen such that it reproduces the analytical value (up to numerical precision) when $p=0$. The SDP optimization was implemented by using PICOS \cite{picos} and the package ncpol2sdpa \cite{Wittek2015}, with the solver SDPA-DD \cite{nakata2010numerical} setting a precision of $10^{-12}$. The dashed lines are the lower bounds for the min-entropy of the non-sequential version of the protocol at NPA order 4.}
    \label{fig:scantheta}
\end{figure}

In Fig.\ \ref{fig:scantheta} we show the lower bound on the min-entropy as a function of the parameter $\theta$ for several values of depolarizing noise $p$. 
The lower bound is obtained by leveraging semidefinite programming (SDP) techniques combined with the Navascués-Pironio-Acin (NPA) hierarchy \cite{navascues_bounding_2007,navascues_convergent_2008}.
Specifically, the NPA order is set to $2+AAB+ABB$, which in the case $p=0$ is sufficient to reproduce the analytical result up to numerical precision and the SDP optimization is performed by using PICOS \cite{picos} and utility functions of ncpol2sdpa \cite{Wittek2015}.
The used solver is SDPA-DD \cite{nakata2010numerical} with a numerical precision of $10^{-12}$.
The constraints imposed to the SDP optimization are the ideal correlations mixed with completely random ones $P^{exp}_{AB_1B_2} = (1-p) P^{ideal}_{AB_1B_2} + p P^{random}_{AB_1B_2}$.
The figure demonstrates that in the presence of this noise, asymptotic values of $\theta$ are no longer optimal for randomness; instead, the optimal value occurs at an intermediate range.
This behavior is likely due to the fact that, for 
$\theta=0,\pi/4$, the sequential protocols considered in this work effectively reduce to a non-sequential one, leading to a discontinuous drop in the certified entropy. For this reason, points close to 
$\theta=0,\pi/4$ are more sensitive to noise than intermediate points.
This phenomenon aligns with observations reported in \cite{Padovan2024}.
For comparison, we also report the lower bound on the min-entropy of the original non-sequential protocol, estimated at NPA order 4. We observe that, even at $p=10^{-2}$, there remain points where the sequential scheme still provides a slight advantage. 
More generally, we recall that the performance of the non-sequential protocol can always be recovered by setting $\theta=0,\pi/4$.

\section{Conclusions}
We have proposed a class of simple quantum sequential protocols, relying on measurements of Pauli observables on a maximally entangled qubit state and designed for device-independent cryptographic tasks.

Under device-independent and i.i.d.\ assumptions, we have analytically demonstrated that the ideal and perfectly known sequential correlations arising in this scenario can serve cryptographic purposes, as these correlations are shown to be extremal within the set of sequential quantum correlations, and extremality ensures that the optimal strategy for any potential eavesdropper attempting to guess the random bits is limited to betting on the most probable measurement outcomes.
We leave the study of the robustness against noise and detection efficiency as future work, which can already be pursued by leveraging numerical methods \cite{navascues_bounding_2007,navascues_convergent_2008, Brown2024,chung2025generalizednumericalframeworkimproved,masini2024}.
Moreover, although we considered a scenario with three parties choosing between two possible dichotomic measurements, more complex scenarios can be explored. 
For example, the number of sequential parties could be increased, or the number of inputs could be expanded by considering self-testing protocols that involve all Pauli matrices \cite{Bowles2018} and by extending them sequentially. 
In that case, the three users could achieve 3 bits of global randomness, independently of the strength parameter of Bob$_1$.
Additionally, we note that the maximally entangled state is not necessary for applying our sequential construction.
One may further generalize our results by considering non-maximally entangled states, exploiting the self-testing properties of some tilted Bell inequality \cite{Bamps2015}.
Another natural generalization would be to relax the i.i.d. assumption. This could be achieved by carefully applying and, if necessary, modifying techniques such as the Entropy Accumulation Theorem \cite{Dupuis2020, Metger2022} or Quantum Probability Estimation \cite{Zhang2020}.
Furthermore, memory effects between experimental rounds could invalidate the core physical assumptions of our sequential scenario, namely the strict requirement that the first Bob's measurement remains uninfluenced by the second Bob's subsequent choices. Therefore, a careful analysis of these effects is necessary.
Finally, in practical implementations one is not limited to extracting randomness from a single designated input sequence. Although many protocols privilege a particular ``generation'' setting by assigning lower sampling probabilities to the other inputs used for self-testing, randomness can in principle be extracted from multiple input sequences.
As a simple illustration, consider the sequential CHSH scenario of Example~1. In the ideal case, Bob's local guessing probability, taking into account all possible input sequences, can be written as
$
G(B|E) = p_{1}/2 + p_{00}\cos^2(\theta)/2 + p_{01}/4,
$
where the coefficients $p_{1}$, $p_{00}$, and $p_{01}$ denote the sampling probabilities associated with the different input sequences, as indicated by their subscripts.
This quantity is always larger than or equal to the non-sequential CHSH guessing probability value of $1/2$, showing even in this case an advantage from the sequential scenario. In practical implementations, however, noise must also be taken into account. In particular, noise affects different input--output configurations in distinct ways and therefore impacts their respective entropy contributions. A proper optimization must balance these effects together with the choice of input probabilities, ensuring that the weighted combination of sequences provides a genuine advantage under realistic, noisy conditions.

We emphasize that the device-independent security of our sequential protocols explicitly relies on the sequential ordering between the two Bobs. 
Consequently, a proper device-independent implementation must not only address known loopholes \cite{Larsson2014, Cabello2007} but also ensure that Bob$_1$ receives his measurement outcomes before Bob$_2$'s choice of input. 
This requirement excludes pure photonic implementations with a tree-like sequential measurement structure, such as the ones in \cite{Foletto2020, Foletto2021, Padovan2024}, where the whole outcomes history is retrieved with a photon detector placed at Bob$_2$'s location. 
Although pure photonic implementations could leverage quantum non-demolition measurements of photon states, they remain inefficient \cite{Pryde2004,Pryde2005}.
Nevertheless, while photons may not be the ideal candidates for these protocols at present, alternative implementations based on matter-based quantum states and measurements could be explored \cite{Storz2023, Krutyanskiy2023, Nadlinger2022, Zhou2024, Redeker2023, vanLeent2022}.

The key feature of the proposed sequential scheme is the inclusion of a non-projective measurement, which can be viewed as a weak version of an observable, and the subsequent measurement of the \textit{same} observable. 
It is well-established that quantum non-projective measurements exhibit a trade-off between information gain and state disturbance \cite{Silva2015}: stronger measurements provide more information but induce greater disturbance on the measured state. 
Accordingly with \cite{Padovan2024}, our results show that even when the measurement strength approaches zero, yielding arbitrarily small information gain, the measurement outcomes still remain random in a device-independent setting.
The results of this work can open avenues for future research on device-independent randomness generation and key distribution protocols.
For example, we note a connection between the sequential scenario considered in this work and the routed Bell experiments \cite{chaturvedi2023, Lobo2024, roydeloison2024}, in which sequentiality is present between the switch station and the two possible subsequent Bobs.
Could the methods adopted here further improve the performance of protocols based on routed Bell experiments?
In addition to their application in device-independent security protocols, sequential quantum schemes are also valuable for investigating fundamental aspects of quantum mechanics, such as contextuality \cite{Budroni2022,Cabello2008,Cabello2016,Guhne2010,Kumari2023}.
Our findings may be useful for advancing research in this area as well.
For example, they could help in finding new inequalities whose violation certifies contextuality in the same way that Bell inequalities certify nonlocality.

\section*{Acknowledgments}
We thank Giuseppe Vallone and Giulio Foletto for the useful comments and suggestions.
This work was supported by European Union's Horizon Europe research and innovation program under the project Quantum Secure Networks Partnership (QSNP), grant agreement No 101114043. 
Views and opinions expressed are however those of the authors only and do not necessarily reflect those of the European Union or European Commission-EU. 
Neither the European Union nor the granting authority can be held responsible for them.

\bibliographystyle{quantum}
\bibliography{biblio_formatted.bib}

\newpage
\appendix

\section{Extremality and security of the sequential quantum correlations}
\label{app:extremality}
In this appendix we recall the device-independent sequential scenario, the definition of extremal correlations, and the connection between extremality and the security condition defined in the main text.
Finally, we show that when security condition holds, the correlations certify all the randomness obtainable from the observed outcomes.

\subsection{Projective and commuting framework for sequential measurements}
\label{app:proj_comm_framework}
In this section we summarize the projective framework adopted in this work.
Useful references are \cite{Bowles2020, Padovan2024, Gallego_2014}.

For simplicity, we only consider a scenario involving two sequential users, Bob$_1$ and Bob$_2$. 
The generalization with an arbitrary number of sequential user can be found in \cite{Padovan2024}.

\paragraph{Correlations in terms of mixed state and Positive Operator Valued Measurements}
Quantum Mechanics prescribes that the evolution of a quantum state is in general governed by the action of Kraus operators and measurements are generally described in terms of POVMs.
Specifically, if Bob$_1$ measures an initial state $\rho$ with a POVM $\{{\bobkraus{y_1}{b_1}}^\dagger \bobkraus{y_1}{b_1}\}_{b_1}$, satisfying the normalization condition
\begin{equation}
    \sum_{b_1} {\bobkraus{y_1}{b_1}}^\dagger \bobkraus{y_1}{b_1} = \openone\,, \qquad \forall y_1\,,
\end{equation}
then the post-measurement state conditioned on the output $b_1$ is $\bobkraus{y_1}{b_1}\,\rho\, {\bobkraus{y_1}{b_1}}^\dagger$, which is normalized to $P_{B_1}(b_1|y_1)$.

Subsequently, this post-measurement state is sent to Bob$_2$ which perform a second POVM $\qty{F^{y_2}_{b_2}}_{b_2}$.
The joint correlations between Bob$_1$ and Bob$_2$ are then computed as
\begin{equation}
    P_{B_1B_2}(b_1,b_2|y_1,y_2) = \Tr[F^{y_2}_{b_2}\til \bobkraus{y_1}{b_1}\,\rho\, {\bobkraus{y_1}{b_1}}^\dagger] = \Tr[{\bobkraus{y_1}{b_1}}^\dagger\,F^{y_2}_{b_2}\, \bobkraus{y_1}{b_1}\,\rho]\,,
\end{equation}
where we used the cyclicity of the trace in the last equality.
In particular, we note that the joint correlations can be written in terms of a combined POVM $\qty{{\bobkraus{y_1}{b_1}}^\dagger\,F^{y_2}_{b_2}\, \bobkraus{y_1}{b_1}}_{b_1,b_2}$, satisfying the normalization condition
\begin{equation}
    \sum_{b_1,b_2} {\bobkraus{y_1}{b_1}}^\dagger\,F^{y_2}_{b_2}\, \bobkraus{y_1}{b_1} = \sum_{b_1} {\bobkraus{y_1}{b_1}}^\dagger \bobkraus{y_1}{b_1} = \openone\,, \qquad \forall y_1,y_2\,,
\end{equation}
and sequential constraints
\begin{equation}
    \sum_{b_2} {\bobkraus{y_1}{b_1}}^\dagger\,F^{y_2}_{b_2}\, \bobkraus{y_1}{b_1} = \sum_{b_2} {\bobkraus{y_1}{b_1}}^\dagger\,F^{y_2'}_{b_2}\,\bobkraus{y_1}{b_1} = {\bobkraus{y_1}{b_1}}^\dagger\bobkraus{y_1}{b_1}\,,  \qquad \forall y_2,y_2'\,.
\end{equation}
Physically, sequential constraints state that the actions of Bob$_2$ do not affect the actions of Bob$_1$.
As a consequence, sequential constraints hold also in the correlations:
\begin{equation}
    \sum_{b_2} P_{B_1B_2}(b_1,b_2|y_1,y_2) = \sum_{b_2} P_{B_1B_2}(b_1,b_2|y_1,y_2') = P_{B_1}(b_1|y_1)\,.
\end{equation}

\paragraph{From POVMs and mixed state to commuting projectors and pure state}
Thanks to Stinespring/Naimark dilation, we can always find an Hilbert space such that the correlations obtained with the POVM $\qty{{\bobkraus{y_1}{b_1}}^\dagger\,F^{y_2}_{b_2}\, \bobkraus{y_1}{b_1}}_{b_1,b_2}$ can be written in terms of a PVM $\qty{\bobseq{y_1y_2}{b_1b_2}}_{b_1,b_2}$ \cite{Bowles2020}:
\begin{equation}
    P_{B_1B_2}(b_1,b_2|y_1,y_2) = \Tr[{\bobkraus{y_1}{b_1}}^\dagger\,F^{y_2}_{b_2}\, \bobkraus{y_1}{b_1}\,\rho] = \expval{\bobseq{y_1y_2}{b_1b_2}}{\psi}\,,
\end{equation}
where $\ket{\psi}$ is the purification of the initial state $\rho$ equipped also with the additional degrees of freedom for the dilation of the POVM.
The PVM measurement satisfy normalization, projectivity and sequentiality:
\begin{align}
    \sum_{b_1,b_2} \bobseq{y_1y_2}{b_1b_2} &= \openone \qquad \forall y_1,y_2 \\
    \bobseq{y_1y_2}{b_1b_2}\bobseq{y_1y_2}{b_1'b_2'} &= \delta_{b_1,b_1'}\delta_{b_2,b_2'} \bobseq{y_1y_2}{b_1b_2}\qquad \forall y_1,y_2 \label{eq:proj_cond_bowl}\\
    \sum_{b_2}  \bobseq{y_1y_2}{b_1b_2} &= \sum_{b_2}  \bobseq{y_1y_2'}{b_1b_2} \qquad \forall y_1,y_2
\end{align}

The last equation allows for the definition of projector operators independent on $y_2$, describing the marginal probabilities of Bob$_1$:
\begin{align}
    \bobproj{y_1}{b_1} &\equiv \sum_{b_2}  \bobseq{y_1y_2}{b_1b_2}\,,\\
    \bobproj{y_1}{b_1}\bobproj{y_1}{b_1'} &= \delta_{b_1,b_1'}\bobproj{y_1}{b_1}\,,\\
    P_{B_1}(b_1|y_1) &= \expval{\bobproj{y_1}{b_1}}{\psi}\,, 
\end{align}
where the second equation can be verified using \eqref{eq:proj_cond_bowl}.

Analogously, we can define projective operators describing the marginal probabilities of Bob$_2$ that in general depend also on $y_1$:
\begin{align}
    \bobproj{y_1y_2}{b_2} &\equiv \sum_{b_1}  \bobseq{y_1y_2}{b_1b_2}\,,\\
    \bobproj{y_1y_2}{b_2}\bobproj{y_1y_2}{b_2'} &= \delta_{b_2,b_2'}\bobproj{y_1y_2}{b_2}\,,\\
    P_{B_2}(b_2|y_1,y_2) &= \expval{\bobproj{y_1y_2}{b_2}}{\psi}\,. 
\end{align}
Moreover, since $\bobproj{y_1}{b_1}$ and $\bobproj{y_1y_2}{b_2}$ can be written in terms of the same projectors $ \bobseq{y_1y_2}{b_1b_2}$, they commute:
\begin{equation}
    \qty[\bobproj{y_1}{b_1},\bobproj{y_1y_2}{b_2}] = 0\,.
\end{equation}
Finally, their product reproduce the joint correlations between the two users:
\begin{equation}
    \bobproj{y_1}{b_1}\bobproj{y_1y_2}{b_2} = \bobseq{y_1y_2}{b_1b_2}\,.
\end{equation}

\subsection{Relation between extremality and security}
We utilize the projective framework reviewed in the main text and in the previous section, and presented in \cite{Bowles2020,Padovan2024}.
In particular, we consider a pure quantum state $\ket{\psi}\in \mathcal{H}_A\otimes\mathcal{H}_B\otimes\mathcal{H}_E$ and some sets of projective measurements, $\qty{\aliceproj{x}{a}}$ and $\qty{\bobproj{y_1}{b_1}, \bobproj{y_1,y_2}{b_2}}$, acting on $\mathcal{H}_A$ and $\mathcal{H}_B$ respectively.
We will refer to the set $\qty{\ket{\psi}, \qty{\aliceproj{x}{a}}, \qty{\bobproj{y_1}{b_1}, \bobproj{y_1,y_2}{b_2}}}$ as a quantum realization of the correlations.
The Alice-Bob$_1$-Bob$_2$ correlations are retrieved through the Born rule by tracing out over Eve's space $\mathcal{H}_E$:
\begin{equation}
\label{eq:AB1B2corrs}
    P_{AB_1B_2}(a,\mathbf{b}|x,\mathbf{y}) = \expval{\aliceproj{x}{a}\otimes\bobproj{y_1}{b_1}\bobproj{y_1,y_2}{b_2}\otimes \openone_E}{\psi} = \Tr_{AB}\qty[\aliceproj{x}{a}\otimes\bobproj{y_1}{b_1}\bobproj{y_1,y_2}{b_2} \til \rho_{AB}]\,,
\end{equation}
where we defined $\rho_{AB}\equiv \Tr_E[\dyad{\psi}]$.
The operators in the above expression satisfy the orthogonality and normalization constraints:
\begin{equation}
\begin{split}
    \aliceproj{x}{a}\aliceproj{x}{a'} &= \delta_{aa'}\aliceproj{x}{a} \ , \\
    \bobproj{y_1}{b_1}\bobproj{y_1}{b_1'} &= \delta_{b_1b_1'}\bobproj{y_1}{b_1} \ , \\
    \bobproj{y_1y_2}{b_2}\bobproj{y_1y_2}{b_2'} &= \delta_{b_2b_2'}\bobproj{y_1y_2}{b_2}  \ , \\
    \sum_a \aliceproj{x}{a} &= \openone_A  \ , \\
    \sum_{b_1} \bobproj{y_1}{b_1} &= \sum_{b_2} \bobproj{y_1y_2}{b_2} = \openone_B\,.
\end{split}
\end{equation}
Moreover, from the sequential constraints, the following commutation holds
\begin{equation}
    \left[\bobproj{y_1}{b_1}, \bobproj{y_1y_2}{b_2}\right] = 0\,.
\end{equation}
The operations that Eve can perform in her own space can be assumed without loss of generality to be some set of projectors $\qty{E_e}$ satisfying $\sum_e E_e = \openone_E$.
Therefore, the correlations \eqref{eq:AB1B2corrs} can be written as a convex sum of some set of (possibly different) correlations parameterized by the index $e$:
\begin{equation}
\begin{split}
\label{eq:convexcombi}
    P_{AB_1B_2}(a,\mathbf{b}|x,\mathbf{y}) &= \sum_e \Tr_{AB}\qty[\aliceproj{x}{a}\otimes\bobproj{y_1}{b_1}\bobproj{y_1,y_2}{b_2} \til \Tr_E\qty[E_e \dyad{\psi}]] \\
    &= \sum_e P_E(e) \Tr_{AB}\qty[\aliceproj{x}{a}\otimes\bobproj{y_1}{b_1}\bobproj{y_1,y_2}{b_2} \til \rho_{AB}^e] \\
    &= \sum_e P_E(e)  P_{AB_1B_2}(a,\mathbf{b}|x,\mathbf{y},e)\,,
\end{split}
\end{equation}
where we defined the normalized states $\rho_{AB}^e \equiv \Tr_E[E_e \dyad{\psi}] / P_E(e)$ and $P_E(e) \equiv \Tr_E[E_e \Tr_{AB}[\dyad{\psi}]]$.
We say that the correlations $P_{AB_1B_2}(a,\mathbf{b}|x,\mathbf{y})$ are extremal if they cannot be written as a convex combination of other different correlations.
In other terms, if $P_{AB_1B_2}(a,\mathbf{b}|x,\mathbf{y})$ are extremal and \eqref{eq:convexcombi} holds, then it must be $P_{AB_1B_2}(a,\mathbf{b}|x,\mathbf{y},e) = P_{AB_1B_2}(a,\mathbf{b}|x,\mathbf{y})$ for each $e$.
Since the aim of Eve is to guess the outcomes of the three users by leveraging the knowledge of the final convex decomposition in \eqref{eq:convexcombi}, extremality guarantees that Eve cannot have more information about the experimental correlations than the information that Alice, Bob$_1$ and Bob$_2$ have.

As shown in \cite{Franz2011}, the extremality of correlations is in one to one correspondence with the security of correlations. 
For the sake of clarity, we adapt the discussion of \cite{Franz2011} to our sequential scenario.
A probability distribution $P_{AB_1B_2}(a,\mathbf{b}|x,\mathbf{y})$ is called secure if it does not factorize, $P_{AB_1B_2} \neq P_A(a|x)P_{B_1B_2}(\mathbf{b}|\mathbf{y})$, and if for any quantum realization and any operator $E$ it holds
\begin{equation}
\label{eq:securitydef}
    \expval{\aliceproj{x}{a}\otimes\bobproj{y_1}{b_1}\bobproj{y_1,y_2}{b_2}\otimes E}{\psi} =  \Tr[E \dyad{\psi}] P_{AB_1B_2}(a,\mathbf{b}|x,\mathbf{y})\,.
\end{equation}
With this definition, it is possible to show that a probability distribution $P_{AB_1B_2}(a,\mathbf{b}|x,\mathbf{y})$ is secure if and only if it is extremal in $\mathcal{Q}_{SEQ}\setminus \mathcal{C}$, with $\mathcal{C}$ the set of correlations that can be written as convex combination of results of local deterministic measurements \cite{Franz2011}.
Note that, since signaling is present from Bob$_1$ to Bob$_2$, we consider them as a unique Bob with four inputs and four outputs, whose measurements are defined in terms of the projectors $\bobseq{y_1y_2}{b_1b_2}\equiv \bobproj{y_1}{b_1}\bobproj{y_1y_2}{b_2}$ satisfying sequentiality constraints.
See also the previous section and \cite{Bowles2020} for further details.

We highlight that, in \cite{Franz2011}, the authors imposed only commutation between operators of separated parties, rather than the tensor product structure of the entire Hilbert space.
This is not an issue for our proofs, since we can always redefine the local operators to act on the entire Hilbert space as  $\aliceproj{x}{a} \equiv \aliceproj{x}{a}\otimes\openone_B\otimes\openone_E$, and similarly for Bob's and Eve's operators, thereby realizing a strategy with commuting operators.
In fact, although we specifically consider tensor product realizations, the following demonstrations can be expressed solely by invoking the commutation between Alice, Bob, and Eve. 
Therefore, they also apply to the more general case of commuting realizations.

\subsection{Randomness from security}
We now show that security guarantees that all the randomness in the observed correlations is certified.
From this point forward, we denote the set of chosen inputs used to generate randomness as $\mathbf{r} \equiv (x^\star, y_1^\star, y_2^\star)$.
In particular, we show that the worst-case quantum conditional min-entropy, $H_{min}(AB|E)$ reduces to $-\log_2 \max_{ab_1b_2}P_{AB_1B_2}(a,\mathbf{b}|\mathbf{r})$, while the worst-case conditional von Neumann entropy, $H(AB|E)$, reduces to the Shannon entropy $-\sum_{ab_1b_2}P_{AB_1B_2}(a,\mathbf{b}|\mathbf{r})\log_2\qty[P_{AB_1B_2}(a,\mathbf{b}|\mathbf{r})]$.
These two quantities are calculated on the classical-quantum post-measurement state defined as
\begin{equation}
    \rho_{ABE}^{post} \equiv \sum_{a,\mathbf{b}} P_{AB_1B_2}(a,\mathbf{b}|\mathbf{r}) \dyad{a,\mathbf{b}}\otimes \tau_E^{a,\mathbf{b}} \quad \in \mathcal{H}_{A'}\otimes\mathcal{H}_{B'}\otimes\mathcal{H}_{E}\,,
\end{equation}
where $\mathcal{H}_{A'}$ and $\mathcal{H}_{B'}$ are two Hilbert spaces in which Alice and Bob register their outcomes, $\dyad{a,\mathbf{b}}=\dyad{a}\otimes\dyad{\mathbf{b}}$, $\sum_a \dyad{a}=\openone_{A'}$, $\sum_\mathbf{b} \dyad{\mathbf{b}}=\openone_{B'}$ and 
\begin{equation}
    \tau_E^{a,\mathbf{b}} \equiv \frac{\Tr_{AB}\qty[\aliceproj{x^\star}{a}\otimes\bobproj{y_1^\star}{b_1}\bobproj{y_1^\star,y_2^\star}{b_2}\otimes\openone_E\til\dyad{\psi}]}{P_{AB_1B_2}(a,\mathbf{b}|\mathbf{r})} \quad \in \mathcal{H}_{E}\,.
\end{equation}
The statement regarding the two entropies is proved by showing that security implies that the post-measurement state is in product form with Eve's space.
Eve's outcomes are distributed according 
\begin{equation}
    P_E(e|a,\mathbf{b}) = \Tr_E\qty[E_e \til \tau_E^{a,\mathbf{b}}] = \frac{P_{AB_1B_2}(a,\mathbf{b},e|\mathbf{r})}{P_{AB_1B_2}(a,\mathbf{b}|\mathbf{r})} = \frac{\expval{\aliceproj{x}{a}\otimes\bobproj{y_1}{b_1}\bobproj{y_1,y_2}{b_2}\otimes E_e}{\psi}}{P_{AB_1B_2}(a,\mathbf{b}|\mathbf{r})}\,,
\end{equation}
and imposing \eqref{eq:securitydef} we obtain $P_E(e|a,\mathbf{b}) = P_E(e)$, indicating that Eve's outcomes do not depend on the ones of Alice and the Bobs.
Since we are not fixing a particular quantum realization, and \eqref{eq:securitydef} holds for any Eve's measurement, the states $\tau_E^{a,\mathbf{b}}$ and $\Tr_{AB}[\dyad{\psi}]$ are equal.
Therefore, the post-measurement state reduces to
\begin{equation}
    \rho_{ABE}^{post} = \qty[\sum_{a,\mathbf{b}} P_{AB_1B_2}(a,\mathbf{b}|\mathbf{r}) \dyad{a,\mathbf{b}}]\otimes \Tr_{AB}[\dyad{\psi}] \equiv \rho^{post}_{AB}\otimes\rho^{post}_E\,.
\end{equation}
We can then directly conclude about the conditional von Neumann entropy.
Indeed, by definition $H(AB|E)_{\rho} = H(\rho^{post}_{ABE}) - H(\rho^{post}_{E})$ and, since the $\rho^{post}_{ABE}$ has a product form, for additivity we obtain that $H(AB|E)_{\rho} = H(\rho^{post}_{AB})$.
Finally, since $\rho^{post}_{AB}$ has a diagonal form, its von Neumann entropy is the Shannon entropy of $P_{AB_1B_2}(a,\mathbf{b}|\mathbf{r})$.

For the min-entropy we calculate the guessing probability as \cite{renner2006,Konig2009,Brown2020}
\begin{equation}
\begin{split}
    G(AB|E) &\equiv \max_{\qty{E_{a\mathbf{b}}}} \sum_{a,\mathbf{b}} P_{AB_1B_2}(a,\mathbf{b}|\mathbf{r}) \Tr_E[E_{a\mathbf{b}} \tau_{E}^{a,\mathbf{b}}] \\
    &= \sum_{a,\mathbf{b}} P_{AB_1B_2}(a,\mathbf{b}|\mathbf{r}) \max_{\qty{E_{a\mathbf{b}}}} \Tr_E[E_{a\mathbf{b}} \rho^{post}_E] \\
    &= \sum_{a,\mathbf{b}} P_{AB_1B_2}(a,\mathbf{b}|\mathbf{r})P_{E}(a,\mathbf{b}) \leq \max_{a,\mathbf{b}}  P_{AB_1B_2}(a,\mathbf{b}|\mathbf{r})\,.
\end{split}
\end{equation}
Finally, $H_{min}(AB|E) = -\log_2 G(AB|E) = -\log_2 \max_{a,\mathbf{b}}  P_{AB_1B_2}(a,\mathbf{b}|\mathbf{r})$.

\section{Proof of security of the sequential correlations}
\label{app:proofs}
In this appendix, we propose a specific quantum realization of the sequential correlations using projective measurements.
Such realization is obtained from the one in Sec.\ \ref{sec:seq_prot} through a Stinespring dilation of the POVMs.
Furthermore, we recall in detail the expression of the Tsirelson bound and, using its maximal saturation, we prove the security of the correlations.

\subsection{Sequential correlations with projective measurements}
As introduced in the main text, sequential quantum correlations can always be written in terms of a pure state and projective measurements.
Here we show such state and measurements for the sequential protocols proposed in the main text.

The state is 
\begin{equation}
    \ket{\psi} = \ket{\phi^+}_{AB}\Bigl[\cos\theta \ket{+}_{B'} + \sin\theta\ket{-}_{B'}\Bigr] \quad \in \mathcal{H}_{A}\otimes\mathcal{H}_{B}\otimes\mathcal{H}_{B'} = \mathbb{C}^2\otimes\mathbb{C}^2\otimes\mathbb{C}^2\,,
\end{equation}
where the kets $\qty{\ket{\pm}}$ are the eigenstates of $\sigma_x$.

Alice's measurements remain the same introduced in the main text:
\begin{align}
    \aliceobs{0} &= \cos \alpha_0 \sigma_x + \sin \alpha_0 \sigma_z \label{eq:A0_app} \,,  \\ 
    \aliceobs{1} &= \cos \alpha_1 \sigma_x + \sin \alpha_1 \sigma_z \label{eq:A1_app} \,,
\end{align}
and act on $\mathcal{H}_A$. 
On Bob's side, the measured observables are
\begin{align}
    \bobobs{0} &= \sigma_x \otimes \sigma_x \ , \\
    \bobobs{1} &= \Bigl[\cos \beta_1 \sigma_x + \sin \beta_1 \sigma_z \Bigr] \otimes \openone_{B'} \ , \\
    \bobobs{00} &= \sigma_x \otimes \openone_{B'} \ , \\
    \bobobs{01} &= \cos\delta \sigma_x\otimes\openone_{B'} + \sin\delta \sigma_z\otimes\sigma_z \,
\end{align}
and act on $\mathcal{H}_B\otimes \mathcal{H}_{B'}$.
The previous measurements can be identified as projective dilation of the protocol in Sec.\ \ref{sec:seq_prot} by using the same  procedure used in \cite{Padovan2024} (see supplementary material therein), up to replacing the unitary evolution in \cite{Padovan2024} with the CNOT gate
\begin{equation}
    U = \dyad{+}\otimes\openone_{B'} + \dyad{-}\otimes\sigma_z
\end{equation}
acting on $\mathcal{H}_{B}\otimes\mathcal{H}_{B'}$.

\subsection{Tsirelson bound}
\label{app:tsirelson}
The Tsirelson bound utilized in the main text is based on self-testing assumptions for specific measurements within the protocol. 
Once the problem is mapped to a standard bipartite scenario through Stinespring dilation, as detailed in Appendix \ref{app:proj_comm_framework}, these assumptions imply
\begin{equation}
\label{eq:selftest_eqs}
\begin{split}
	\ket{\psi} &\uli \ket{\phi^+}_{AB}\ket{\xi}_{B'E}  \ , \\
	\aliceobs{0}\ket{\psi} &\uli \Bigl[\cos\alpha_0 \sigma_x + \sin\alpha_0 \sigma_z \Bigr]\otimes \openone_B \otimes \openone_{B'E}\til\ket{\phi^+}_{AB}\ket{\xi}_{B'E} \ , \\
	\aliceobs{1}\ket{\psi} &\uli \Bigl[\cos\alpha_1 \sigma_x + \sin\alpha_1 \sigma_z\Bigr]\otimes \openone_B\otimes \openone_{B'E}\til \ket{\phi^+}_{AB}\ket{\xi}_{B'E} \ , \\
	\bobobs{1}\ket{\psi} &\uli \openone_A \otimes \Bigl[\cos\beta_1 \sigma_x + \sin\beta_1 \sigma_z \Bigr]\otimes \openone_{B'E}\til  \ket{\phi^+}_{AB}\ket{\xi}_{B'E} \ , \\
	\bobobs{00}\ket{\psi} &\uli \openone_A \otimes \sigma_x\otimes \openone_{B'E}\til \ket{\phi^+}_{AB}\ket{\xi}_{B'E}\,,
\end{split}
\end{equation}
where the symbol $\uli$ indicates an equality up to local isometries.
In particular, on Alice's side we can reconstruct two unitary operators associated to her $\sigma_x$ and $\sigma_z$ in the ideal protocol:
\begin{equation}
	\mqty( X_A \\ Z_A ) \equiv \mqty(\cos\alpha_0 & \sin\alpha_0 \\ \cos\alpha_1 & \sin\alpha_1)^{-1}\cdot \mqty( \aliceobs{0} \\ \aliceobs{1} )\,.
\end{equation}
As shown in \cite{Wang_2016}, these operators are unitary and anti-commute on the support of the state:
\begin{align}
    X_A^\dagger X_A \ket{\psi} = Z_A^\dagger Z_A \ket{\psi} &= \ket{\psi} \ , \\
    \qty{X_A,Z_A}\ket{\psi} &= 0 \label{eq:alice_anticomm}\,.
\end{align}
From now on, for simplicity, we will omit the tensor product symbol $\otimes$. 
However, we recall that $X_A$ and $Z_A$ commute with any of Bob's and Eve's operators, as they act on different spaces.

At first, we recall that, by construction,
\begin{align}
\label{eq:Bobcommute}
    \qty[\bobobs{0},\bobobs{00}] =  \qty[\bobobs{0},\bobobs{01}] = 0\,.
\end{align}
The Tsirelson bound is then expressed as a function of the Bell operator
\begin{equation}
\label{eq:Std_app}
    S_{\theta,\delta} \equiv \frac{1}{2}\Bigl[\openone - X_A \Bigl( c_1\bobobs{01}+c_2\bobobs{0} \Bigr) - Z_A\Bigl( c_3\bobobs{01}+c_4\bobobs{0} \Bigr)\Bigr]\,.
\end{equation}
This Bell operator can be easily retrieved with the method described in \cite{coccia2025}.
In particular, the coefficients are defined as
\begin{equation}
\label{eq:C_coefficients}
\begin{aligned}
	c_{1} &\equiv \frac{\cos(\delta)\sin^2(2\theta)}{1 - \cos^2(\delta)\cos^2(2\theta)} \ , \\
    c_{2} &\equiv \frac{\cos(2\theta)\sin^2(\delta)}{1 - \cos^2(\delta)\cos^2(2\theta)} \ , \\
    c_{3} &\equiv \frac{\sin(\delta)\sin(2\theta)}{1 - \cos^2(\delta)\cos^2(2\theta)} \ , \\
    c_{4} &\equiv -\frac{\sin(\delta)\cos(\delta)\cos(2\theta)\sin(2\theta)}{1 - \cos^2(\delta)\cos^2(2\theta)}\,,
\end{aligned}
\end{equation}
satisfying 
\begin{align}
    c_1^2+c_2^2+c_3^2+c_4^2 &= 1 \ , \label{eq:c_cond1}\\
    c_1 c_2 + c_3 c_4 &= 0 \label{eq:c_cond}\,.
\end{align}
Under the self-testing assumption, we can state that
\begin{equation}
\label{eq:Ssquared}
    S_{\theta,\delta}^\dagger S_{\theta,\delta} \ket{\psi} = S_{\theta,\delta}\ket{\psi}\,,
\end{equation}
and therefore $\expval{S_{\theta,\delta}}\geq 0$.
Indeed, by direct substitution, we see that
\begin{align}
    S_{\theta,\delta}^\dagger S_{\theta,\delta} \ket{\psi} &= \frac{1}{4}\Bigl[2 S_{\theta,\delta} + \qty{Z_A,X_A}\qty(c_1 \bobobs{01} + c_2 \bobobs{0})\qty(c_3 \bobobs{01} + c_4 \bobobs{0}) + \qty(c_1c_2 + c_3c_4)\qty{\bobobs{0},\bobobs{01}}\Bigr]\ket{\psi}\\
    &= S_{\theta,\delta} \ket{\psi}\,,
\end{align}
where we explicitly used \eqref{eq:alice_anticomm}, \eqref{eq:Bobcommute}, \eqref{eq:c_cond1} and \eqref{eq:c_cond}.
Given \eqref{eq:Std_app} and \eqref{eq:Ssquared}, we deduce that a necessary condition that states and measurements must satisfy in order to saturate $\expval{S_{\theta,\delta}}= 0$ is
\begin{equation}
\label{eq:S_max_violation}
    \ket{\psi} = X_A \qty(c_1 \bobobs{01} + c_2 \bobobs{0})\ket{\psi} + Z_A \qty(c_3 \bobobs{01} + c_4 \bobobs{0}) \ket{\psi} \,,
\end{equation}
indeed, $\expval{S_{\theta,\delta}} = \expval*{S_{\theta,\delta}^\dagger S_{\theta,\delta}} = \norm{S_{\theta,\delta}\ket{\psi}}^2  = 0\implies S_{\theta,\delta}\ket{\psi} = 0$.
By direct calculation, it is possible to verify that such condition is verified by the state and measurements defined in the previous section.

\subsection{Security}
\label{app:securitycalc}
In this section we prove the separability of the correlations Alice-Bobs-Eve with respect to Eve, as described in \eqref{eq:securitydef}.

All the correlations $P_{AB_1B_2}(a,\mathbf{b}|x,\mathbf{y})$ can be retrieved from the expectation values of the operators:
\begin{center}
\begin{tabular}{c|cc}
\hspace{1cm} \textit{Self-tested operators} \hspace{1cm} & \hspace{1cm} \textit{Correlations to be certified} \hspace{1cm} &\\
\toprule
$Z_A$ & $\expval{\bobobs{0}}$ \\
$X_A$ & $\expval{\bobobs{01}}$ \\
$\bobobs{1}$ & $\expval{\bobobs{0}\bobobs{00}}$ \\
$\bobobs{00}$ & $\expval{\bobobs{0}\bobobs{01}}$ \\
$Z_A\bobobs{1}$ & $\expval{Z_A\bobobs{0}}$ \\
$X_A\bobobs{1}$ & $\expval{X_A\bobobs{0}}$ \\
$Z_A\bobobs{00}$ & $\expval{Z_A\bobobs{01}}$ \\
$X_A\bobobs{00}$ & $\expval{X_A\bobobs{01}}$ \\
& $\expval{Z_A\bobobs{0}\bobobs{00}}$ \\
& $\expval{Z_A\bobobs{0}\bobobs{01}}$ \\
& $\expval{X_A\bobobs{0}\bobobs{00}}$ \\
& $\expval{X_A\bobobs{0}\bobobs{01}}$ \\
\end{tabular}
\end{center}
Indeed, we can write
\begin{equation}
\begin{split}
    P_{AB_1B_2}(a,\mathbf{b}|x,\mathbf{y}) &= \expval{\aliceproj{x}{a}\otimes \bobproj{y_1}{b_1} \bobproj{y_1y_2}{b_2}} \\
    &= \expval{\frac{\openone + a \aliceobs{x}}{2}\otimes \frac{\openone + b_1 \bobobs{y_1}}{2}\frac{\openone + b_2 \bobobs{y_1y_2}}{2}} \\
    &= \frac{1}{8} \Bigl[ 1 + a \expval{\aliceobs{x}} + b_1 \expval{\bobobs{y_1}} + b_2 \expval{\bobobs{y_1y_2}} + a b_1 \expval{\aliceobs{x}\bobobs{y_1}} + a b_2 \expval{\aliceobs{x}\bobobs{y_1y_2}} \\
    &\hspace{0.8cm} + b_1b_2\expval{\bobobs{y_1}\bobobs{y_1y_2}}  + a b_1b_2\expval{\aliceobs{x}\bobobs{y_1}\bobobs{y_1y_2}} \Bigr]\,.
\end{split}
\end{equation}

The aim is to prove that, for any operator $O$ in the above two columns, it holds $\expval{O E} = \expval{O}\expval{E}$ for any Hermitian operator $E$ of Eve.
Given the self-test \eqref{eq:selftest_eqs}, this holds true already for the operators in the left column.
Moreover, it holds also
\begin{equation}
\label{eq:ZAXABE}
    \expval{Z_A X_A \bobobs{} E} \in i \mathbb{R}
\end{equation}
for any Bobs' Hermitian operator $\bobobs{}$.
Indeed, thanks to the self-testing:
\begin{align}
    Z_A \ket{\psi} &\uli \sigma_z \otimes \openone_{B}\otimes \openone_{B'E} \ket{\phi^+}_{AB}\ket{\xi}_{B'E}\,, \\
    X_A \ket{\psi} &\uli \sigma_x \otimes \openone_{B}\otimes \openone_{B'E} \ket{\phi^+}_{AB}\ket{\xi}_{B'E}\,,
\end{align}
and therefore
\begin{equation}
     \expval{Z_A X_A \bobobs{} E} = \qty(\bra{\psi}Z_A)\qty(X_A \bobobs{} E \ket{\psi}) = \bra{\xi}\bra{\phi^+}\sigma_z \sigma_x \otimes B_{BB'} \otimes E \ket{\phi^+}\ket{\xi} = i \expval{\sigma_y \otimes B_{BB'} \otimes E}\,,
\end{equation}
where $B_{BB'}$ results from the action of the local isometry on $B$, and $\expval{\sigma_y \otimes B_{BB'} \otimes E}\in \mathbb{R}$ since $B_{BB'}$ and $E$ are Hermitian.

To enhance clarity, we present the proof's calculations as a bulleted list.
\begin{itemize}
\item \emph{$\bobobs{0}$ and $\bobobs{01}$:}
    Define
    \begin{align}
        B_x &= c_1 \bobobs{01} + c_2 \bobobs{0} \ , \\
        B_z &= c_3 \bobobs{01} + c_4 \bobobs{0} \ .
    \end{align}
    If the matrix made of the $c_i$ is invertible (which is true for $\delta \neq 0$ and $\theta \neq k \frac{\pi}{4}$), then $\bobobs{0}$ and $\bobobs{01}$ can be written in terms of $\bobobs{x}$ and $\bobobs{z}$. 
    From the boundary condition \eqref{eq:S_max_violation}, applying $X_A E$ to both sides by and considering the mean values, we obtain
    \begin{equation}
      \expval{X_A E} = \expval{\bobobs{x} E} + \expval{X_A Z_A \bobobs{z} E} = 0\,
    \end{equation}
    where we used the self-testing conditions on Alice's operator to set the entire expression to $\expval{X_A E} = \expval{X_A}\expval{E} = 0$. Using that the mean value of  $\expval{Z_A X_A \bobobs{} E}$ is purely imaginary, Eq.\ \eqref{eq:ZAXABE}, we can conclude that $\expval{\bobobs{x} E}=0$. 
    Similarly, applying $Z_A E$ to \eqref{eq:S_max_violation},  we can also derive the condition $\expval{\bobobs{z} E} = 0$. Therefore we conclude that
    \begin{align}
        \expval{\bobobs{0} E} &= 0=\expval{\bobobs{0}}\expval{E} \ , \\
       \expval{\bobobs{01} E} &= 0=\expval{\bobobs{01}}\expval{E} \, .
    \end{align}
\item 
    \emph{$Z_A \bobobs{0} \bobobs{01}$ and $X_A \bobobs{0} \bobobs{01}$:}
    Starting again from the boundary condition \eqref{eq:S_max_violation} and applying $\bobobs{0} E$ or $X_A Z_A \bobobs{0} E$ we obtain, respectively
    \begin{align}
    \label{eq:una}
       \expval{B_0 E} &= 0= c_1\expval{X_A B_0 B_{01}E} + c_3 \expval{Z_A B_0 B_{01}E}\, , \\
    \label{eq:dua}
        \expval{X_A Z_A \bobobs{0} E} &= 0 = -c_1\expval{Z_A \bobobs{0} \bobobs{01} E} + c_3 \expval{X_A \bobobs{0} \bobobs{01} E}\,.
    \end{align}
        Considering now the linear combination $c_2 \cdot \eqref{eq:una} + c_4 \cdot \eqref{eq:dua}$, and using the property \eqref{eq:c_cond} we arrive to
    \begin{equation}
      0 = (c_2 c_3 - c_1c_4) \expval{Z_A \bobobs{0} \bobobs{01} E}\,,
    \end{equation}
    and therefore $\expval{Z_A \bobobs{0} \bobobs{01} E} = 0=\expval{Z_A \bobobs{0} \bobobs{01}}\expval{E}$, if $c_2 c_3 - c_1c_4 \neq 0$ (which is true if $\delta\neq 0$ and $\theta \neq k\frac{\pi}{4}$).

Considering instead the combination $c_4 \cdot \eqref{eq:una} - c_2 \cdot \eqref{eq:dua}$, and using again \eqref{eq:c_cond}, we end up with
    \begin{equation}
       0 = (c_1 c_4 - c_2c_3) \expval{X_A \bobobs{0} \bobobs{01} E}\,,
    \end{equation}
    and therefore $\expval{X_A \bobobs{0} \bobobs{01} E} = 0=\expval{X_A \bobobs{0} \bobobs{01}}\expval{E}$ when $\delta\neq 0$ and $\theta \neq k\frac{\pi}{4}$.  
\item 
\emph{$X_A \bobobs{0}\bobobs{00}$ and $Z_A \bobobs{0}\bobobs{00}$:}
    The mean value of these operators is computed by recalling that, from the self-test \eqref{eq:selftest_eqs}, we have
    \begin{equation}
    \label{eq:X_B}
        \bobobs{00}\ket{\psi} = X_A \ket{\psi}\,.
    \end{equation}
    Applying $X_A \bobobs{0}$ to both sides leads to
    \begin{equation}
        \expval{X_A \bobobs{0} \bobobs{00} E} = \expval{X_A \bobobs{0} X_A E} = \expval{\bobobs{0} E} = 0=\expval{\bobobs{0}}\expval{E}\,.
    \end{equation}
    Applying instead  $Z_A \bobobs{0}$ , it holds
    \begin{equation}
        \expval{Z_A \bobobs{0} \bobobs{00} E}= \expval{Z_A X_A \bobobs{0} E} = - \expval{X_A Z_A \bobobs{0} E} 
    \end{equation}
    where we used the anti-commutation of $X_A$ and $Z_A$. However, using the property  \eqref{eq:Bobcommute}, we can also write
    \begin{equation}
      - \expval{X_A Z_A \bobobs{0} E} = \expval{Z_A \bobobs{0} \bobobs{00} E}= \expval{Z_A \bobobs{00} \bobobs{0} E}= \expval{X_A Z_A \bobobs{0} E}\,,
    \end{equation}
    so that
    \begin{equation}
        \expval{Z_A \bobobs{0} \bobobs{00} E} = 0=\expval{Z_A \bobobs{0} \bobobs{00}}\expval{E} \ . 
    \end{equation}
\item 

\emph{All the remaining ones:}
The remaining correlations to consider are:
\begin{align}
    \expval{Z_A \bobobs{01} E} &= c_3 \expval{E} + c_4 \expval{\bobobs{0}  \bobobs{01} E} \ , \\
    \expval{Z_A \bobobs{0} E} &= c_4 \expval{E} + c_3 \expval{\bobobs{0} \bobobs{01}  E} \ , \\
    \expval{X_A \bobobs{01} E} &= c_1 \expval{E} + c_2 \expval{\bobobs{0} \bobobs{01} E} \ , \\
    \expval{X_A\bobobs{0} E} = \expval{\bobobs{0} \bobobs{00} E} &= c_2 \expval{E} + c_1 \expval{\bobobs{0} \bobobs{01} E} \ . 
\end{align}

By using above equations and \eqref{eq:X_B}, we can evaluate $\expval{Z_A B_{00} E}$ from \eqref{eq:S_max_violation}:
\begin{equation}
\begin{split}
    0 = \expval{Z_A B_{00} E} &= c_1 \expval{Z_A X_A B_{00} B_{01}E} + c_2 \expval{Z_A X_A B_{00} B_0 E} + c_3 \expval{B_{00} B_{01}E} + c_4 \expval{ B_{00} B_0 E} \\
    &= - c_1 \expval{Z_A B_{01}E} + c_2 \expval{Z_A B_0 E} + c_3 \expval{X_A B_{01}E} + c_4 \expval{X_A B_0 E} \\
    &= 2 c_2 c_4 \expval{E} + 2 c_2 c_3 \expval{B_0 B_{01} E}
\end{split}
\end{equation}
where in the second line we used $\qty[B_0,B_{00}]=0$.
We then find
\begin{equation}
    \expval{B_0 B_{01} E} = -\frac{c_4}{c_3} \expval{E}\,.
\end{equation}
We note that above calculations allow to conclude that the correlations saturating the Tsirelson bound are unique in the sequential scenario.
Indeed, by fixing $E=\openone_E$ in the previous discussion, we can calculate all the correlations between Alice, Bob$_1$ and Bob$_2$ by imposing only the boundary condition \eqref{eq:S_max_violation} and the sequential constraint \eqref{eq:Bobcommute}.
\end{itemize}

Thanks to the security condition, we can directly calculate the global min and von Neumann entropy of the outcomes of the measurements in the protocol.  
Specifically, we are interested in the input $x^\star$ on Alice's side and the sequence $\mathbf{y}^\star = (0,1)$ on Bob's side, for which we have the following formulas
\begin{equation}
\label{eq:entropies_protocols}
\begin{split}
    H_{min} &= -\log_2\qty[\max_{a,b_1,b_2}P_{AB_1B_2}(a,b_1,b_2|x^\star,y_1^\star,y_2^\star)] = 3 - \log_2\qty[\max_{a,b_1,b_2} f_{x^\star,y_1^\star,y_2^\star}(a,b_1,b_2)] \ , \\
    H &= - \sum_{a_1,b_1,b_2} P_{AB_1B_2}(a,b_1,b_2|x^\star,y_1^\star,y_2^\star) \log_2\Bigl[P_{AB_1B_2}(a,b_1,b_2|x^\star,y_1^\star,y_2^\star)\Bigr]  \\
    &=3 - \frac{1}{8}\sum_{a_1,b_2,b_2} f_{x^\star,y_1^\star,y_2^\star}(a,b_1,b_2)\log_2[f_{x^\star,y_1^\star,y_2^\star}(a,b_1,b_2)]\,,
\end{split}
\end{equation}
with
\begin{equation}
    f_{x^\star,y_1^\star,y_2^\star}(a,b_1,b_2) = 1 + b_1b_2\cos\delta \cos 2\theta + a\cos\alpha_{x^\star} (b_1\cos2\theta + b_2\cos\delta) + ab_2\sin\alpha_{x^\star}\sin2\theta\sin\delta \,,
\end{equation}
where the parameters $\delta$ and $\theta$ on Bob's side can be chosen to maximize the randomness, together with the parameter $\alpha_{x^\star}$ characterizing Alice's input $x^\star$.
\end{document}